\documentclass[reprint,secnumaRabic,amssymb,nobibnotes,aps,pra]{revtex4-2}

\usepackage{graphicx}
\usepackage{dcolumn}
\usepackage{bm}
\usepackage{amsmath}

\newcommand{\bra}[1]{\langle #1 \vert}
\newcommand{\ket}[1]{\vert #1 \rangle}

\begin{document}

\title{
Rydberg state engineering: A comparison of tuning schemes for continuous frequency sensing}

\author{Samuel Berweger} 
\author{Nikunjkumar Prajapati}
\author{Alexandra B. Artusio-Glimpse}
\author{Andrew P. Rotunno}
\author{Roger Brown}
\author{Christopher L. Holloway}
\author{Matthew T. Simons}
\affiliation{National Institute of Standards and Technology, Boulder, CO, 80305}%

\author{Eric Imhof}
\author{Steven R. Jefferts}
\affiliation{Northrop Grumman, Woodland Hills, CA, 91361}

\author{Baran N. Kayim}
\author{Michael A. Viray}
\author{Robert Wyllie}
\author{Brian C. Sawyer}
\affiliation{Georgia Tech Research Institute, Atlanta, GA, 30332}

\author{Thad G. Walker}
\affiliation{Department of Physics, University of Madison-Wisconsin, Madison, WI, 53706}

\date{\today}

\begin{abstract}
On-resonance Rydberg atom-based radio-frequency (RF) electric field sensing methods remain limited by the narrow frequency signal detection bands available by resonant transitions.
The use of an additional RF tuner field to dress or shift a target Rydberg state can be used to return a detuned signal field to resonance and thus dramatically extend the frequency range available for resonant sensing.
Here we investigate three distinct tuning level schemes based on adjacent Rydberg transitions, which are shown to have distinct characteristics and can be controlled with mechanisms based on the tuning field frequency or field strength.
We further show that a two-photon Raman feature can be used as an effective tuning mechanism separate from conventional Autler-Townes splitting. 
We compare our tuning schemes to AC Stark effect-based broadband RF field sensing and show that although the sensitivity is diminished as we tune away from a resonant state, it nevertheless can be used in configurations where there is a low density of Rydberg states, which would result in a weak AC Stark effect.
\end{abstract}

\maketitle

\section*{Introduction}

Rydberg atom sensors have emerged as a promising quantum sensing technology for detecting electric (E) fields in the MHz-GHz range \cite{aly_2022}.
The E fields couple to a resonant transition between two Rydberg states \cite{ostwalder1999}, and the resulting effect on one of these states is measured via the electromagnetically induced transparency (EIT) detected using a two-photon ladder scheme that couples to this state \cite{shaffer2012,holloway2014,fan2015}.
In contrast to traditional antennas, the atom sensor size can be independent of the RF wavelength used and the atoms themselves do not significantly perturb propagating fields.
The full realization of the potential of Rydberg atoms for radio frequency (RF) field sensing for communications and data transfer applications will require simultaneous high data rate bandwidth \cite{susanne2021,song2019,holloway2019,sapiro2020}, sensitivity \cite{prajapati2021,jing2020,gordon2019,kumar2017}, and frequency multiplexing.

Rydberg E-field sensing is typically performed using a mixer configuration, where an RF-frequency local oscillator (LO) is used to generate a known intermediate frequency (IF) from the signal, which dramatically improves sensitivity \cite{prajapati2021,jing2020,gordon2019} while simultaneously providing frequency selectivity.
The broadband detection of nonresonant fields is readily possible based on the AC Stark shift \cite{meyer2021,li2022}, albeit at the cost of reduced sensitivity compared to resonant field sensing \cite{meyer2020}.
Although resonant Autler-Townes (AT) splitting-based field sensing can achieve sensitivity as high as $\approx$~5~$\mu$V/m$\cdot \sqrt{\rm Hz}$ \cite{prajapati2021,jing2020,gordon2019}, it is only sensitive to RF fields resonant with discrete dipole-allowed Rydberg transitions that are adjacent to the optically coupled state, and thus of limited utility for wideband sensing and multiplexing.
A recently demonstrated way to improve the spectral range over which resonant Rydberg sensing can be used is to engineer the available Rydberg states -- and thus available transitions -- using an additional ``tuner'' RF field \cite{simons2021}.
By using this tuner field to, e.g., shift a target Rydberg state via AT splitting to return a detuned signal field to resonance, the sensitivity can be improved. 

The continuous range of Rydberg states in terms of the principal and angular momentum quantum numbers \cite{brown2022} can provide a broad manifold of available states for a given signal and tuner frequency of interest. 
As such, a wide range of possible arrangements of the tuner and signal fields -- tuning schemes -- are available, each with characteristic tuning behavior and associated benefits and drawbacks. 
In this work, we compare three possible arrangements of tuner and signal fields, including the one described previously \cite{simons2021}, which we here term \emph{sequential tuning}, as well as two additional schemes termed \emph{split tuning} and \emph{inverse sequential tuning}.
Several of these tuning schemes further allow tuning by several distinct mechanisms, including an EIT feature arising from a two-RF photon Raman transition.

\section*{Experimental}

\begin{figure*}
	\includegraphics[width=0.95\textwidth]{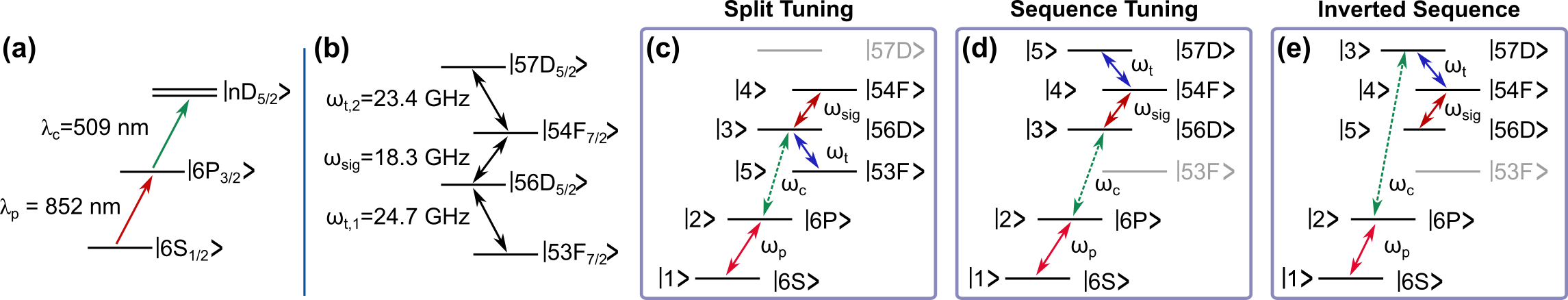}
	\caption{
        (a) Schematic of the Cs EIT ladder scheme to access (b) the Rydberg state manifold used.
        In all cases we use the 56D-54F RF transition for our signal field.
        We explore three different tuning configurations by optically exciting to either the 56D or 57D states and tuning with either the 53F-56D or 54F-57D transitions.
        The state arrangements are shown for our \emph{split tuning} (c), \emph{sequence tuning} (d), and \emph{inverted sequence} (e) schemes.
        State numbers in (c)--(e) correspond to the five-level models described in the Appendix.
	}
	\label{schematic}
\end{figure*}

Our experimental setup is based on the cesium two-photon EIT ladder scheme shown in Fig.~{\ref{schematic}}(a), which has found widespread application in Rydberg electrometry \cite{cox2018,chopinaud2021,jing2020,holloway2017}.
Electric fields are sensed via the Autler-Townes splitting of the Rydberg state by an adjacent RF-frequency transition. 
For E-field detection we use the Rydberg-atom mixer approach introduced in \cite{simons2019}, where we apply an LO that is detuned from the signal (Sig) field of interest by approximately 10~kHz and the resulting atom-mixed EIT beat frequency (beat note) is then demodulated via lock-in detection.
In order to facilitate sweeping of the LO/Sig frequencies while maintaining a constant beat note frequency we use amplitude modulation (AM) of the LO field to simulate the heterodyne/superheterodyne Rydberg mixer configuration.

To enable direct comparison between the different tuning schemes we use the same 56D$_{5/2}$-54F$_{7/2}$ transition for our LO/Signal field in all cases as illustrated in Fig.~\ref{schematic}(b).
This transition manifold is chosen to leverage the high linearity of the D-F transitions \cite{chopinaud2021} and the absence of nearby transition frequencies as is the case for, e.g., the S-P transitions \cite{robinson2021}.
Our tuning schemes are arranged by optically exciting to either the 56D or 57D states and tuning with either the 53F-56D or 54F-57D transitions. 
The corresponding state arrangements for our \emph{split tuning}, \emph{sequence tuning}, and \emph{inverted sequence} schemes are shown in Fig.~\ref{schematic}(c)--(e), respectively.

\begin{figure*}
	\includegraphics[width=0.75\textwidth]{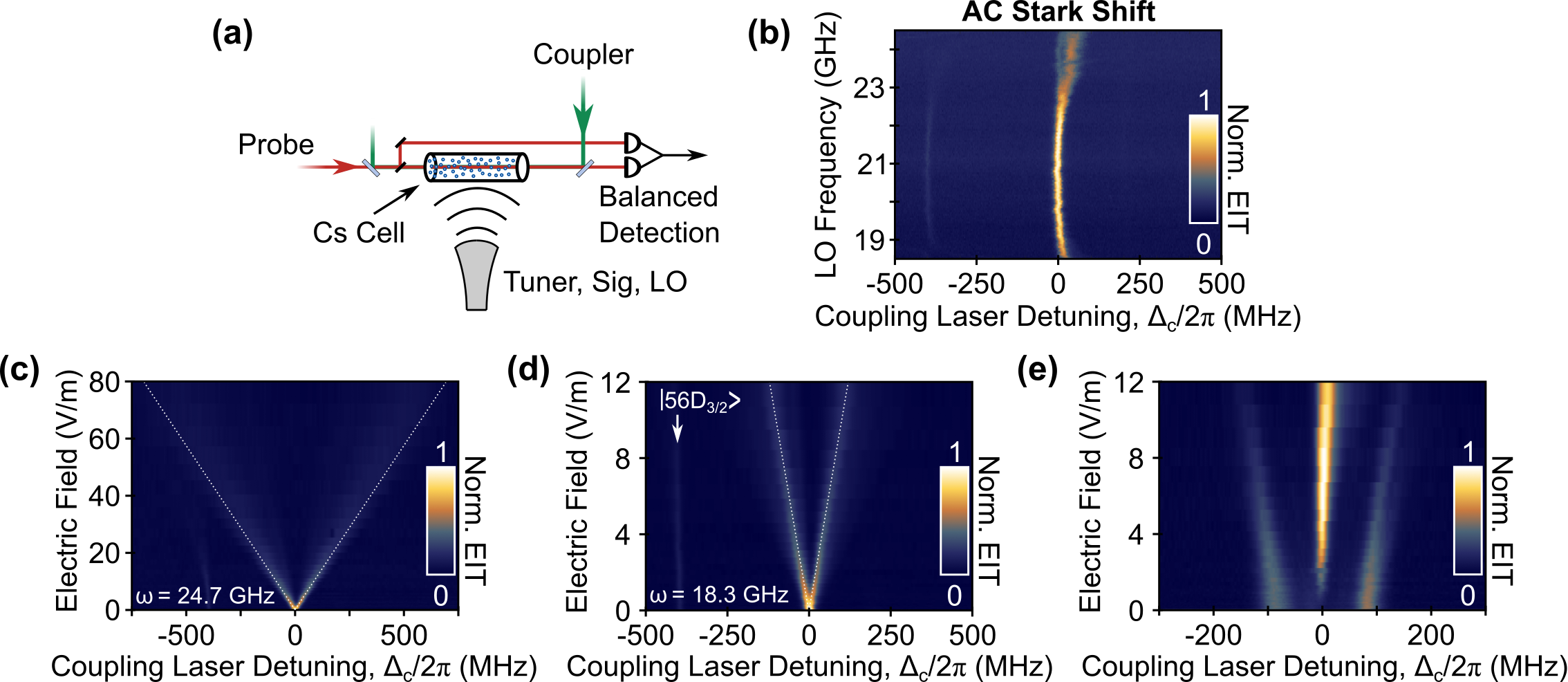}
	\caption{
        (a) Schematic of the experimental setup with counterpropagating probe and coupler beams and the orthogonal horn antenna that is used to broadcast all RF fields used.
        (b) Shift of the 56D EIT peak due to the AC Stark shift induced by the LO field of varying frequency. 
        The resonant 56D-54F and 56D-53F transitions are just out of view below and above the image, respectively.
        (c) AT splitting of the 56D EIT peak with an RF field applied to the 56D-53F tuning transition at 24.7 GHz along with (d) the splitting of the same peak in response to an RF field applied to the 56D-54F signal/LO transition at 18.3 GHz.
        Dashed lines are a guide to the eye.
        (e) With the coupling laser frequency set to the 57D state and a constant RF field of strength $\Omega/2\pi$~=~180~MHz applied to the 57D-54F transition (f = 23.4~GHz), we see the emergence of a two-RF photon Raman peak at $\Delta_c$~=~0 as the strength of the second RF field on the 54D-56F transition at f~=~18.3~GHz is increased.
	}
	\label{setup}
\end{figure*}

The coupler and probe laser beams are sent through a Cs vapor cell in a counterpropagating direction and the EIT probe and reference intensity beams are detected using a balanced photdodiode as illustrated in Fig.~\ref{setup}(a).
The RF fields are applied using a horn antenna covering the frequency range 18~GHz -- 26.5~GHz with all optical and RF fields co-polarized orthogonal to the plane of propagation.
Importantly, our choice of RF transitions allows us to use the same horn for all Sig/LO/tuner fields, ensuring that all fields are co-propagating.

The AC stark shift represents an established path and a baseline for broadband Rydberg atom-based field sensing \cite{meyer2021,li2022}.
Shown in Fig~\ref{setup}(b) is the non-resonant AC Stark shift of the 56D EIT peak as a function of LO frequency. 
The field strength is set to 31~V/m at the resonant 24.7~GHz 56D-53F transition and the output power held constant throughout this sweep, but we expect significant variations in field strength within the vapor cell over the frequency range measured due to the frequency-dependent standing mode profile.

Shown in Fig~\ref{setup}(c) is the resonant AT splitting of the 56D$_{5/2}$ EIT peak due to an RF field applied at the 56D-53F tuning transition at 24.7 GHz.
We use an amplifier to generate large tuner fields $>$~120~V/m to yield Rabi frequencies $\Omega_t/2\pi$~$>$~2~GHz.
Shown in Fig~\ref{setup}(d) is the AT splitting of the same peak due to a field applied on the 56D-54F signal/LO transition at 18.3 GHz.
For the signal/LO transitions we achieve more moderate maximum fields on the order of 15~V/m.
Also shown is the EIT peak due to the 56D$_{3/2}$ transition at $\Delta_c/2\pi$~=~$-396$~MHz that we use to set our laser-scanned frequency axis.

Lastly, in Fig.~\ref{setup}(e) we show the effect of two simultaneously applied RF fields producing a new EIT peak. 
Here, the coupling laser is resonant with the 57D EIT peak and we apply a constant RF field of strength $\Omega/2\pi$~=~180~MHz set to the 57D-54F transition at f~=~23.4~GHz that produces the AT doublet seen.
As we increase the strength of the second RF field at f~=~18.3~GHz on the 54F-56D transition, we see the emergence of a new EIT peak feature near $\Delta_c$~=~0.
This feature is due to a two-RF photon Raman transition between the 57D and 56D states.
We base this conclusion on the observation that we see this peak when $\Delta_t$~=~$-\Delta_{sig}$ for our signal/tuner arrangement, i.e, the sum of the photon energies matches that of the 56D-57D transition (see also the discussion of the inverted sequence scheme below, as well as appendix~\ref{sec:simple3}).
Also, in our previous work we observed a residual EIT peak at $\Delta_c$~=~0 with two resonant RF fields applied, which we attributed to transitions between magnetic sublevels that are forbidden with the $\pi$ transitions induced by our linear polarization \cite{robinson2021}.
However, here our EIT peak structure can be reproduced by our numerically solved 5-level model as well as an analytical 3-level model (see Appendices) without the need for additional magnetic sublevels or angular momentum states.
It is notable that this EIT feature was not previously observed in comparable two-RF photon schemes driven by a single RF field \cite{anderson2014,xue2021}.

\section*{Results}

\begin{figure*}[!htb]
	\includegraphics[width=.95\textwidth]{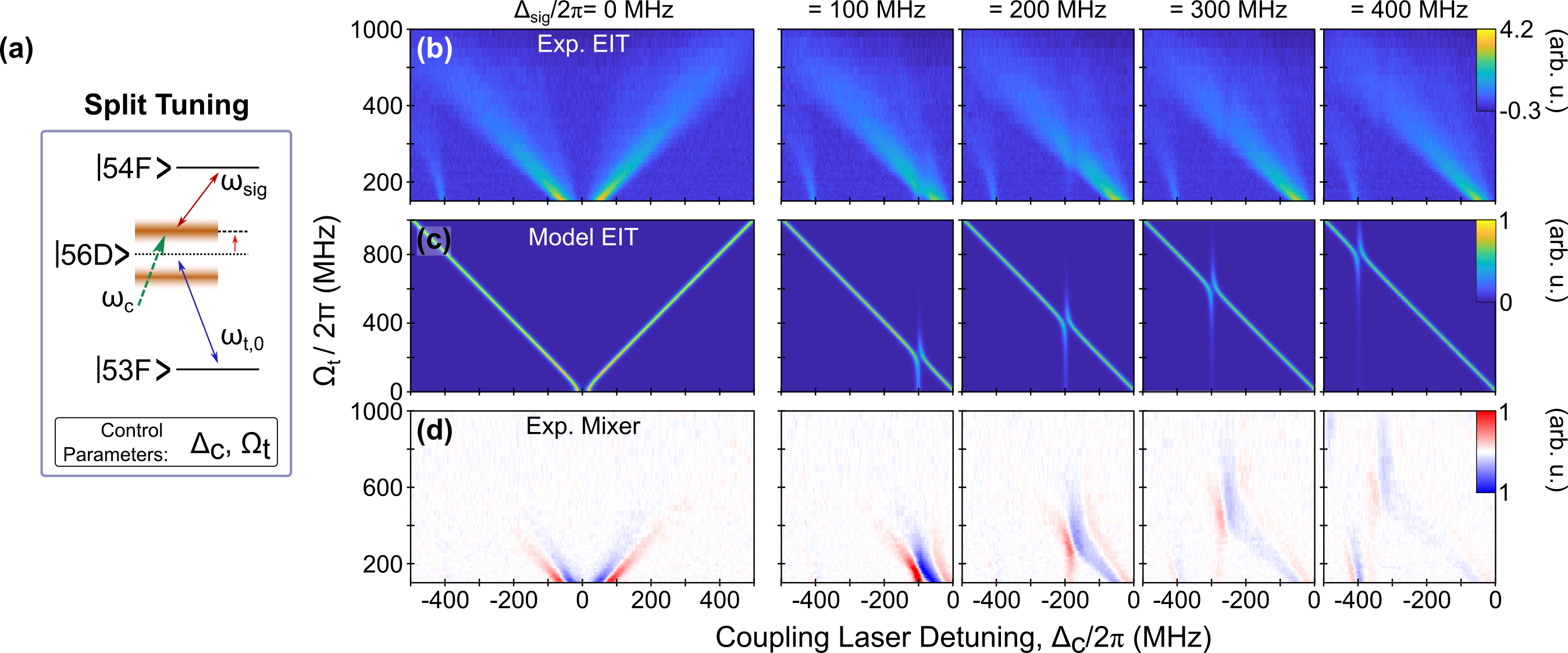}
	\caption{
        (a) Schematic of the states used for power tuning in the split tuning scheme.
        (b) False color plots of the experimental EIT as a function of the tuner Rabi frequency and the coupler laser detuning, with different values of signal frequency detuning for each plot indicated.
        (c) The corresponding modeled EIT shows the location of the tuning peaks obtained, which are clearly visible in (d) the Rydberg mixer plots.
	}
	\label{split}
\end{figure*}

{\bf Split Tuning:} We begin by discussing the conceptually simple split tuning scheme schematically described in Fig.~\ref{split}(a).
We refer to the tuning mechanism presented here for the split tuning scheme as \emph{power tuning} to reflect the fact that we use the strength of the tuner field ($\Omega_t$) as the control parameter to return a signal field for a given $\Delta_{sig}$ to resonance.
Here the signal and tuning fields are both adjacent to the optically coupled state -- 56D in this case.
We use the tuning field to induce AT-splitting the target state (56D) and the Sig/LO fields are returned to resonance when one of the resulting peaks matches the detuned Sig/LO frequencies and coupler frequencies:
\begin{equation}
\Omega_t/2=|\Delta_{sig}| \quad {\rm and} \quad \Delta_{sig}=-\Delta_c \,\,\, .
\label{eq1}
\end{equation}
Shown in Fig.~\ref{split} are the measured Fig.~\ref{split}(b) and modeled Fig.~\ref{split}(c) EIT spectra as a function of $\Omega_c$ with $\Omega_{sig}/2\pi$~=~30~MHz and $\Delta_{sig}$ indicated. The details of the model are presented in Appendix~\ref{sec:mastereq}.
The experimental traces appear similar and are dominated by the tuner-induced AT splitting with residual signal field-induced AT splitting at weak tuner field strengths. 
The modeled data shows an avoided crossing that appears near the condition where $\Delta_{sig}$~=~$-\Delta_c$~=~ $\Omega_t/2$.
These features are also seen in the experimental EIT traces, though they are less well-defined due to peak broadening not present in the model.
These tuning features can be more readily resolved using amplitude modulation as shown in Fig.~\ref{split}(d), which also serves to underscore the practical implications of using a tuning scheme.

For large signal field detuning values $>$150 MHz we see improved benefits of the tuning scheme relative to the residual EIT signal with no applied tuning field or laser detuning.
Although the models clearly show the expected behavior of Eq.~(\ref{eq1}), experimentally we find that $\Delta_{sig}$~$>$~$\Delta_c$.
We attribute this primarily to the presence of the 56D$_{3/2}$ transition, which is clearly visible at $\Delta_c/2\pi$~=~-396~MHz.
Because of the simultaneous presence of both the 56D$_{3/2}$ and the dressed 56D$_{5/2}$ state for $\Delta_c/2\pi$~$\approx$~400~MHz, the resulting mixing shifts the position of the dressed state.
While we show an illustrative tuning range of only 400~MHz here in part due to limitations arising from the presence of the 56D$_{3/2}$ state, we note that arbitrarily large tuning ranges are possible in principle, but they are practically limited by peak broadening at larger tuner field strengths.

\begin{figure*}[!tb]
	\includegraphics[width=.95\textwidth]{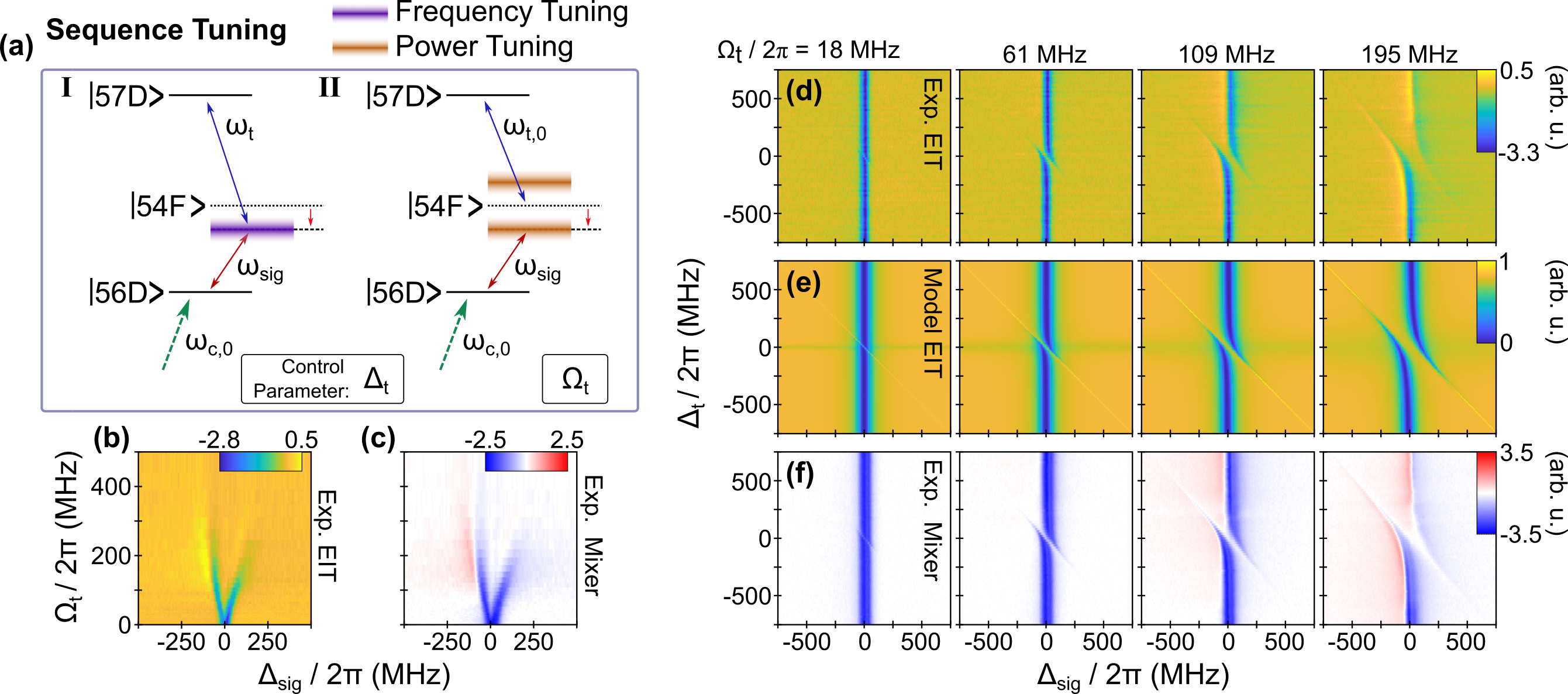}
	\caption{
        (a) Schematic of the tuning schemes used in the sequential arrangement. 
        Experimental EIT (b) and mixer (c) plots of power tuning.
        Here, the signal field is set to the resonant 56D-54F transition and the tuning field is applied to the 54F-57D transition with increasing strength, inducing the observed AT-splitting in the 54F state.
        A set of false-color plots of sweeps of $\Delta_t$ and $\Delta_{sig}$ showing the experimental EIT (d), the modeled EIT (e), and the experimental mixer data (f) as a function of increasing tuner field strength.
	}
	\label{sequence}
\end{figure*}

{\bf Sequential Tuning:} We now turn to the previously considered case of sequential tuning \cite{simons2021}, which can in principle be tuned using either the Raman peak or power tuning as schematically shown in Fig.~\ref{sequence}(a).
The power tuning is shown in the false color plots of the experimental EIT shown in Fig.~\ref{sequence}(b) and the corresponding mixer signal in Fig.~\ref{sequence}(c).
Here the signal field links the optically coupled 56D state to 54F while the tuner field applied along 54F-57D is used to split and tune the energy levels of 54F.
For these plots we lock the coupler laser to the EIT maximum and sweep $\Delta_{sig}$ while increasing $\Omega_t$ and maintaining $\Delta_t$~=~0.
The tuner-induced splitting of the 57F state can clearly be seen providing a means to return the detuned signal field to resonance.

The more general utility of sequential tuning is shown in Fig.~\ref{sequence}(d)-(f), showing the experimental EIT, modeled EIT, and the experimental mixer signals, respectively, as a function of $\Delta_{sig}$ and $\Delta_t$ for tuner Rabi frequencies indicated. 
Here the coupler laser is again locked to the 56D EIT maximum, giving rise to an overall large transparency. 
At low tuner Rabi frequencies the effect of the tuner field is limited, and the EIT behavior is dominated by reduced transparency resulting from the signal field-induced AT splitting near $\Delta_{sig}$~=~0.
As the tuner Rabi frequency is increased, it AT splits the 54F state and the resulting dressed states are seen as $\omega_{sig}$ is swept.
At large values of $\Omega_t$, the $\omega_{sig}$-dependent EIT peaks begin shifting and the mixer signal becomes asymmetric due to the AC stark effect.

One further feature of note is seen in Fig.~\ref{sequence}(d)-(f).
A feature of increased EIT is seen for $\Delta_t$~=~$-\Delta_{sig}$ in the model results, which is not clearly seen in the experimental EIT but is apparent in the mixer signal.
This EIT feature is due is the Raman peak previously shown in Fig.~\ref{setup}(e), which will be discussed in further detail in the next section.
 
Sequential tuning as shown here can operate on one of two mechanisms. 
On one hand, it can readily be power tuned, where we hold $\Delta_t$~=~0 and increase $\Omega_t$ in order to control the amount of induced AT splitting.
However, it also allows for \emph{frequency tuning}, where $\Omega_t$ is kept constant and we use $\Delta_t$ as the control parameter to keep the signal field on a resonance by tracking the $\Omega_t$-dependent signal maxima.

\begin{figure*}[!tb]
	\includegraphics[width=\textwidth]{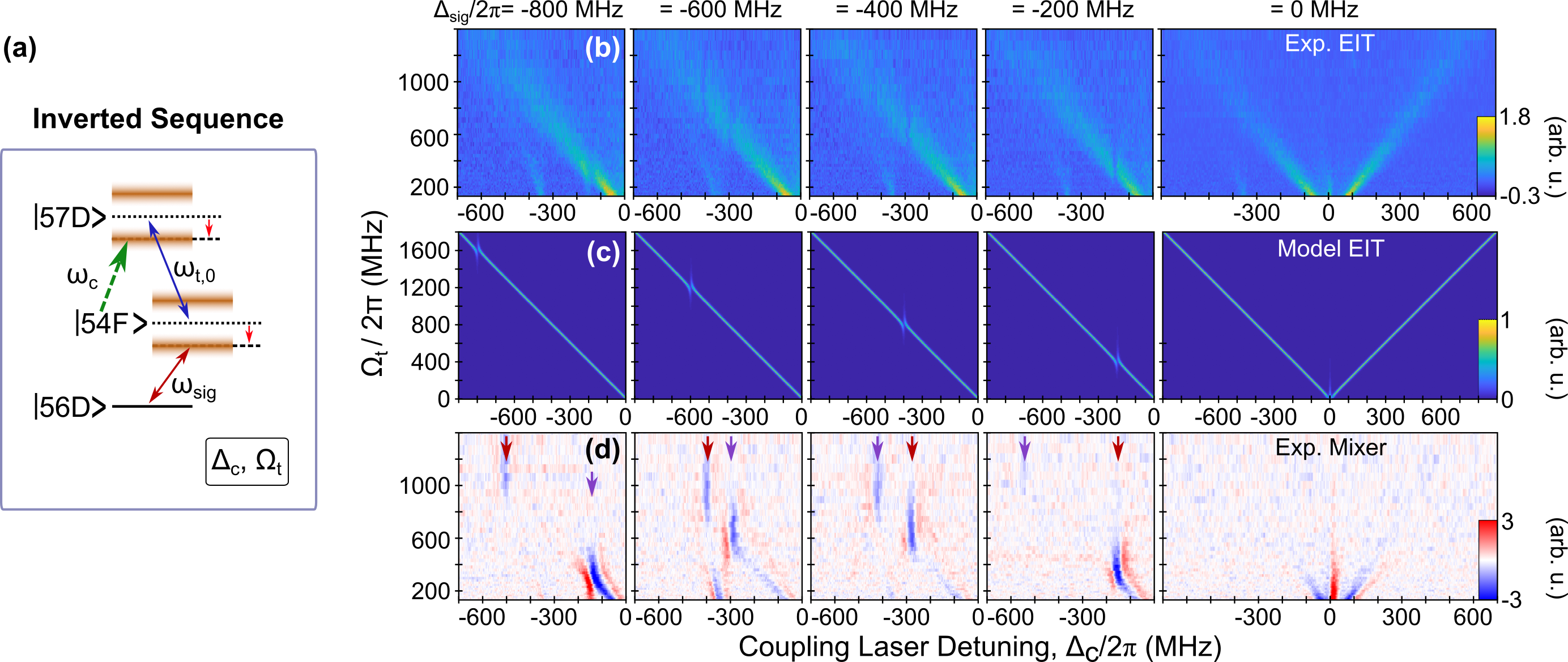}
	\caption{
        (a) Schematic of power tuning in the inverted sequence.
        False color plots of power tuning using the inverted scheme with the experimental raw EIT signal (b), mixer signal (c), and the modeled EIT signal (d).
        Red arrows in (d) indicate EIT features due to the 56D-54F transition while the violet ones indicate those arising from the 57D-55F transition.
        Note different x-axis scaling between experimental and modeled data.
        }
	\label{inverted_pow}
\end{figure*}

{\bf Inverted Sequence:} We now turn to the last of the three tuning schemes addressed in this work, the inverted sequence.
Although conceptually similar to the sequential tuning arrangement, reversing the order of the signal and tuning fields has significant consequences for the tuning mechanisms available and their efficacy.
In order to maintain consistency and enable a direct comparison with the other tuning schemes we maintain the 56D-54F signal transition and the 57D-54F tuner transition as illustrated in Fig.~\ref{inverted_pow}(a).
We invert the sequence by changing the coupler laser frequency to probe the 57D state.
With the tuner transition adjacent to the optically coupled state we can use stronger fields than those desirable for the LO, in order to leverage the benefits of large Rabi rates and associated AT splitting.
On one hand, as further discussed below we can apply large values of $\Omega_t$ in order to effectively split the 57D EIT peak to access the Raman peak feature for frequency tuning.
On the other hand, the inverted scheme also readily lends itself to power tuning comparable to the split scheme discussed above. 

Power tuning of the inverse sequence scheme is shown in Fig.~\ref{inverted_pow}(b)-(d). 
Since we established the baseline utility of power tuning in Fig.~\ref{split}, we show a larger coupling laser frequency range here in order to illustrate a few key features and differences.
First, due to the arrangement of the states used, we find for our inverted scheme here, negative $\Delta_{sig}$ is tuned into resonance at negative values of $\Delta_c$.
Secondly, the Raman peak is seen at $\Delta_c$~=~$\Delta_{sig}$~=~0 in the experimental and modeled EIT, as well as the mixer signal.
Since we show larger values of $\Delta_c$ here, two tuning peaks become visible, where we note that similar peaks would also be seen in the split tuning case.
These arise from two distinct signal transitions: The 56D-54F transition of interest, as well as the 55F-57D transition at $\omega_{sig}$~=~17.4~GHz that is simultaneously measured in a split tuning scheme.
Although our interest here is in the 56D-54F signal transition, the simultaneous presence of an additional signal arising from a split scheme provides a direct comparison.
With a difference in transition dipole moments of $<$~5$\%$, the equivalent signal levels achieved in the peaks at $\Delta_{sig}~=~-600$~MHz and $-400$~MHz reflect the result that the three-level Hamiltonians produce the same eigenvalues for both of these schemes when $\Delta_t~=~0$ (Appendix~\ref{sec:simple3}).
As in the split scheme, in contrast to the models we again find that $|\Delta_{sig}|$~$>$~$\Delta_c$.
We note that this is not generally the case and in the absence of a fine structure peak these two values remain equal in magnitude. 

\begin{figure*}[!tb]
	\includegraphics[width=0.8\textwidth]{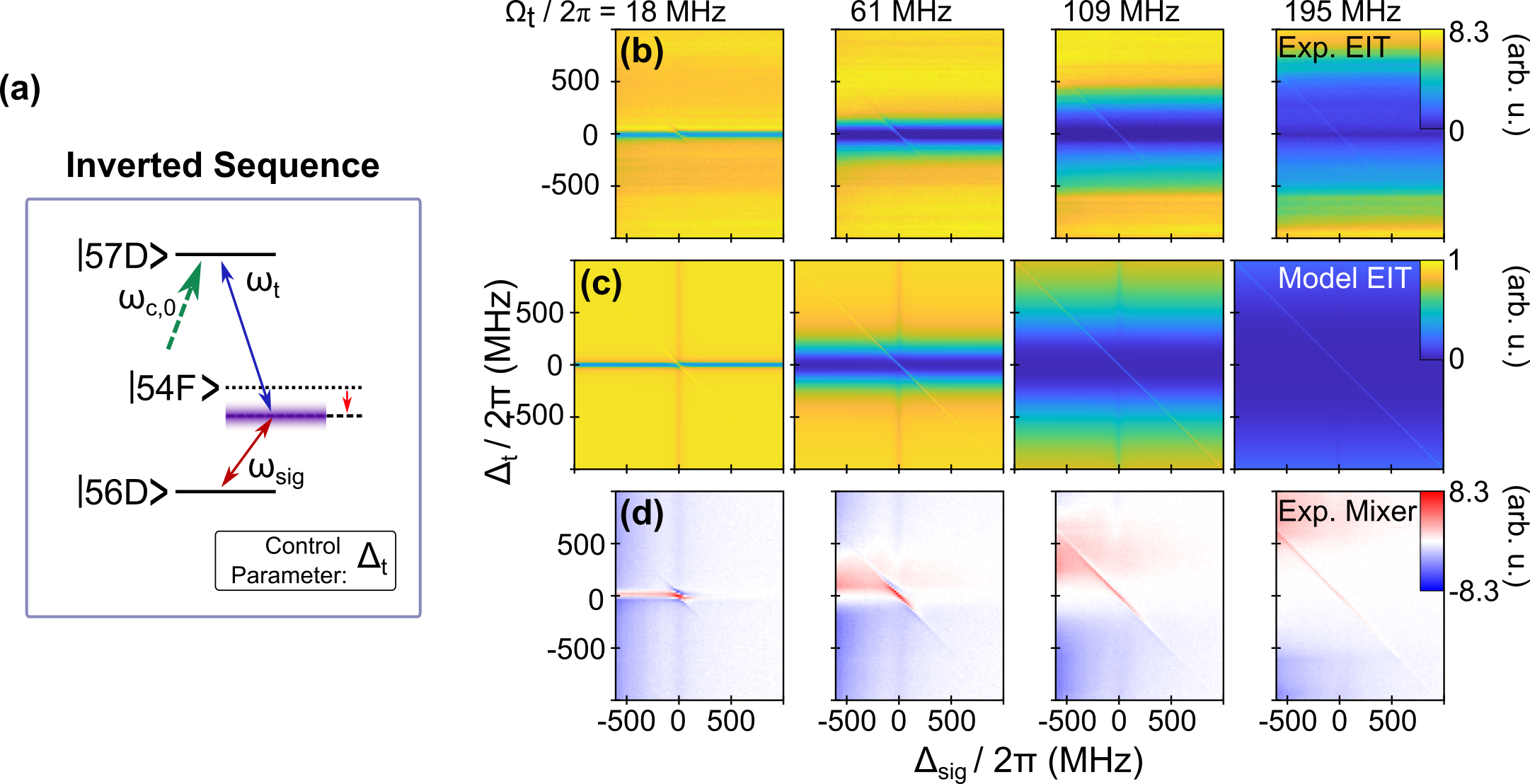}
	\caption{
	(a) Schematic of frequency tuning in the inverted sequence.
    False color plots of frequency tuning using an inverted sequence with the experimental EIT (b), the  modeled EIT (c), and experimental mixer signal (d).
        }
	\label{inverted_freq}
\end{figure*}

The inverted sequence scheme also provides an opportunity to use the Raman peak discussed previously and schematically shown in Fig.~\ref{inverted_freq}(a)
Shown in Fig.~\ref{inverted_freq} are false-color plots of frequency tuning with the coupling laser locked to the EIT maximum.
Although we show a range of $\pm$1~GHz in tuning in the modeled EIT in Fig.~\ref{inverted_freq}(b), the experimental EIT in Fig.~\ref{inverted_freq}(c) and mixer signals in Fig.~\ref{inverted_freq}(d) are shown only for $\Delta_{sig}$~$>$~$-600$~MHz because the signal becomes dominated by the resonant 57D-55F transition at $\Delta_{sig}$~$\approx$~$-900$~MHz.
The residual non-resonant background seen in the mixer signal is due to this transition via the AC Stark effect, which is further influenced by the tuner at higher field strengths.
With the laser locked to the maximum of the 57D EIT peak, the primary effect of the tuner field is AT splitting of the peak, reducing the overall EIT maximum.
At low tuner coupling frequency the induced splitting is small and the magnitude limited for small values of $\Delta_t$.
As $\Omega_t$ increases, however, the values of $\Delta_t$ affected by the AT splitting increases.

The tuning-relevant feature seen in all traces is the increased EIT amplitude on the diagonal where $\Delta_{sig}$~=~$-\Delta_t$, corresponding to the Raman feature seen in Fig.~\ref{setup}(e).
As seen in the model EIT plots, this feature persists over a broad range of frequency detuning largely independent of $\Omega_t$.
However, in practice -- as seen in both the experimental EIT as well as the mixer signal plots -- the tunability of this feature is limited and increases with $\Omega_t$.
This is because this Raman feature is weaker than the resonant EIT peak, and as such only becomes discernible once the EIT peak is sufficiently split to separate it from the Raman peak of interest. 
As a consequence, larger values of $\Delta_{sig}$ require increasingly large values of $\Omega_t$.
The mixer signal clearly reveals the inherent broadband tunability of the frequency tuning in the inverted sequence scheme, but unless the Raman peak is clearly distinguished from the resonant EIT peak, residual AC Stark effects from nearby transitions can interfere with signal detection.
The mixer data at $\Omega_t/2\pi$~=~195~MHz also cautions that the Raman peak-based tuning scheme shown here is approximately of comparable magnitude as the nonresonant background arising from the AC Stark effect.

\subsection*{Sensitivity Comparison}

\begin{figure}[!htb]
	\includegraphics[width=0.3\textwidth]{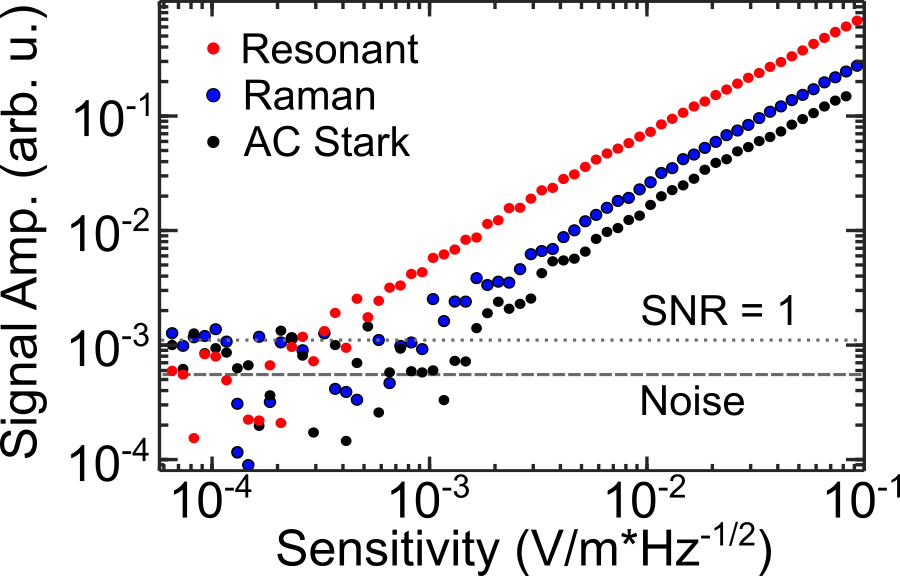}
	\caption{
        Comparison of the baseline sensitivity of on-resonance detection on the 56D-54F transition, the Raman EIT peak, and the AC stark shift at 21.5~GHz.
        }
	\label{sensitivity}
\end{figure}

We provide a direct comparison between tuning schemes by performing baseline sensitivity measurements using a Rydberg mixer using two separately sourced signals detuned by a beat note frequency of 11.2 kHz. 
All measurements are acquired with the coupling laser tuned to the 56D transition used throughout this work.
We use a lock-in time constant of $\tau$~=~1~s, resulting in a bandwidth of BW~=~$1/2\pi \tau$.
Shown in Fig.~\ref{sensitivity} are the sensitivity curves for on-resonant AT-based field sensing, sensitivity using the Raman peak, and off-resonant AC Stark sensing at 21.5~GHz (see Fig.~\ref{setup}b).
Also shown are the measurement noise-floor (dashed line) and the signal amplitude corresponding to a signal-to-noise ratio (SNR) of 1.
As expected, the on-resonant mixer provides the highest sensitivity, approximately 3 times better than the Raman peak, which is in turn twice as sensitive as the AC Stark effect. 
It is important to note that although the AC Stark effect here is measured far-detuned from resonance, it is nevertheless in a frequency range with a large manifold of D-F transitions.
We expect that the relative sensitivity would be lower in a spectral range where fewer nearby transitions are available such as, e.g., at lower principal quantum number n.

\section*{Discussion}

Throughout this work we have described several tuning schemes and mechanisms that can be used to return a detuned signal field to resonance in order to achieve continous frequency detection. 
This is done using an additional tuner RF field in addition to the signal/LO fields used for conventional Rydberg mixer measurements.
These tuning mechanisms rely on either using tuner field-induced AT splitting of a target state to engineer a resonant signal transition (power tuning), or by using a two-RF photon transition to link two real Rydberg states (frequency tuning).
In all cases, there is a net loss in sensitivity compared to the on-resonant sensing case. 
This can be attributed to a distribution of the transition oscillator strength of the bare state into the two AT-split peaks.
Further limitations in tuning sensitivity not accounted for in our models include linewidth broadening due to the  field inhomogeneity typical of vapor cells and the effects of magnetic sublevels \cite{chopinaud2021}.

In all schemes shown here, power tuning leverages tuner field-induced AT splitting.
Thus, an overarching challenge presented by power tuning are the large and uniform tuner fields required by the general condition for an on-resonant tuner field that $\Omega_t$~=~$2*|\Delta_{sig}|$.
Thus, for our chosen states here, a detuning of $\Delta_{sig}/2\pi$~=~1000~MHz requires $\Omega_t/2\pi$~=~2000~MHz, corresponding to an electric field E~$>$~100~V/m.
Although we can readily achieve such fields with our setup, most signal generators (including ours) require an additional amplification stage to produce sufficiently high fields with a horn antenna. 

Leveraging the Raman peak resulting from the coherent interaction of the tuner and LO fields for field sensing presents a new approach here for Rydberg field sensing. 
As we show in Fig.~\ref{sensitivity} the overall sensitivity is not significantly diminished compared to the resonant EIT signal, though we note that at low tuner field strengths, where the resonant EIT peak is not fully split, an additional nonresonant EIT response can contribute too.
In terms of practical implementations, a frequency tuning scheme based on the Raman peak is attractive because it is easier to tune the RF frequencies rather than the laser as required for power tuning.

Lastly, it bears emphasizing that the AC stark sensitivity remains good and benefits from needing merely an LO field rather than an additional tuner field. 
However, the required LO fields are typically significantly stronger than those yielding optimal sensitivity in other schemes.

We conclude by noting that we have demonstrated a set of tuning schemes that can be used for resonant frequency detection by engineering the Rydberg energy levels to return a detuned signal field to resonance using a tuner field. 
These include tuner field-induced AT splitting as well as producing a Raman peak feature that can resonantly link two otherwise dipole-forbidden states.
In the present case the benefits relative to an AC Stark-based approach are marginal.
However, we must emphasize that the details of the sensitivity depend delicately on all aspects of the Rydberg atoms used, including atomic species as well as the principal and angular momentum quantum numbers. 
Our choice of states was driven primarily by the desire to have a manifold of transitions available within the bandwidth op our K-band microwave electronics and horn, which leads to inevitable tradeoffs in terms of nearby transition frequencies.
As such, we do not expect that our results are quantitatively universal, but our experience and modeling does suggest that these tuning schemes are generally applicable.

\appendix
\section{Master-equation model}\label{sec:mastereq}

We use a master-equation model of the EIT signals for the various atomic transition schemes used here. 
Fig.~\ref{schematic}(c)--(e) labels each of the five states addressed in the split, sequence, and inverted sequence tuning configurations for ease in referencing. 
The power of the probe beam measured on the detector (the EIT signal, i.e., the probe transmission through the vapor cell) is given by \cite{berman2011book}
\begin{equation}
P=P_0 \exp\left(-\frac{2\pi L \,\,{\rm Im}\left[\chi\right]}{\lambda_p}\right)=P_0 \exp\left(-\alpha L\right) \,\,\, ,
\label{intensity}
\end{equation}
where $P_0$ is the power of the probe beam at the input of the cell, $L$ is the length of the cell, $\lambda_p$ is the wavelength of the probe laser,  $\chi$ is the susceptibility of the medium seen by the probe laser, and $\alpha=2\pi{\rm Im}\left[\chi\right]/\lambda_p$ is Beer's absorption coefficient for the probe laser.  The susceptibility for the probe laser is related to the density matrix component ($\rho_{21}$)  by the following \cite{berman2011book}
\begin{equation}
\chi=\frac{2\,{\cal{N}}_0\wp_{12}}{E_p\epsilon_0} \rho_{21_D} =\frac{2\,{\cal{N}}_0}{\epsilon_0\hbar}\frac{(d\, e\, a_0)^2}{\Omega_p} \rho_{21_D}\,\,\, ,
\label{chi1}
\end{equation}
where $d=2.02$ \cite{SteckCsData} is the normalized transition-dipole moment for the probe laser, $\Omega_p$ is the Rabi frequency for the probe laser in units of rad/s, and $e$ and $\hbar$ are the elementary charge and reduced Planck's constant, respectively. 
The subscript $D$ on $\rho_{21}$ presents a Doppler averaged value. ${\cal{N}}_0$ is the total density of atoms in the cell and is given by
\begin{equation}
{\cal{N}}_0= \frac{p}{k_B T} \,\, ,
\label{nn}
\end{equation}
where $k_B$ is the Boltzmann constant, $T$ is temperature in Kelvin, and the pressure $p$ (in units of Pa) is given by \cite{SteckCsData}
\begin{equation}
p=10^{9.717-\frac{3999}{T}} 
\label{ppp}
\end{equation}
In eq. (\ref{chi1}), $\wp_{12}$ is the transition-dipole moment for the $\ket{1}$-$\ket{2}$ transition, $\epsilon_0$ is the vacuum permittivity, and $E_p$ is the amplitude of the probe laser E-field.

The density matrix component ($\rho_{21}$) is obtained from the master equation \cite{berman2011book}
\begin{equation}
\dot{\boldsymbol{\rho}}=\frac{\partial \boldsymbol{\rho}}{\partial t}=-\frac{i}{\hbar}\left[\mathbf{H},\boldsymbol{\rho}\right]+\boldsymbol{\cal{L}} \,\,\, ,
\label{me}
\end{equation}
where $\mathbf{H}$ is the Hamiltonian of the atomic system under consideration and ${\boldsymbol{\cal{L}}}$ is the Lindblad operator that accounts for the decay processes in the atom. The $\mathbf{H}$ and $\boldsymbol{\cal{L}}$ matrices for the three different tuning schemes are given below.

\begin{figure*}[t]
	\includegraphics[width=0.7\textwidth]{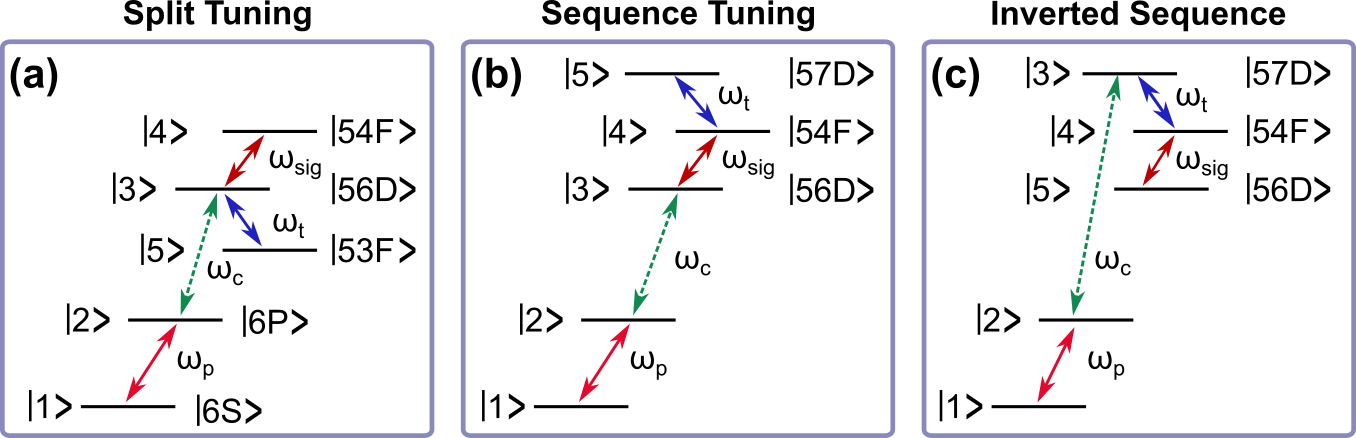}
	\caption{
        Level numbers referenced in the Master-equation model of each tuning scheme.
        }
	\label{fig:numberedLevels}
\end{figure*}

We numerically solve these equations to find the steady-state solution for $\rho_{21}$ for various values of Rabi frequency ($\Omega_i$) and detunings ($\Delta_i$). This is done by forming a matrix with the system of equations for $\dot{\rho}_{ij}=0$. The null-space of the resulting system matrix is the steady-state solution.  The steady-state solution for $\rho_{21}$ is then Doppler averaged~\cite{berman2011book}
\begin{equation}
\rho_{21_D}=\frac{1}{\sqrt{\pi}\,\, u}\int_{-3u}^{3u}\rho_{21}\left(\Delta'_p,\Delta'_c\right)\,\,e^{\frac{-v^2}{u^2}}\,\,dv\,\,\, ,
\label{doppler}
\end{equation}
where $u=\sqrt{2k_B T/m}$ and $m$ is the mass of the atom. We use the case where the probe and coupling laser are counter-propagating. Thus, the frequency seen by the atom moving toward the probe beam is upshifted by $2\pi v/\lambda_p$ (where $v$ is the velocity of the atoms), while the frequency of the coupling beam seen by the same atom is downshifted by $2\pi v/\lambda_c$.  The probe and coupling beam detuning is modified by the following
\begin{equation}
\Delta'_p=\Delta_p-\frac{2\pi}{\lambda_p}v \,\,\,{\rm and}\,\,\,
\Delta'_c=\Delta_c+\frac{2\pi}{\lambda_c}v \,\,\, .
\label{doppler2}
\end{equation}

\subsection{Split Tuning}

For the split tuning scheme shown in Fig.~\ref{schematic}(c), the Hamiltonian can be expressed as:
\begin{equation}
\mathbf{H}=\frac{\hbar}{2}\begin{bmatrix}
    0 & \Omega_p & 0 & 0&0\\
    \Omega_p^* & A & \Omega_c & 0&0\\
    0 & \Omega_c^* & B & \Omega_{sig}&\Omega_{t}\\
    0 & 0 & \Omega_{sig}^* & C & 0 \\
    0 & 0 &\Omega_{t}^* & 0 & D 
\end{bmatrix}\,\, ,
\label{H_split}
\end{equation}
where $\Omega_p$, $\Omega_c$, $\Omega_{sig}$, and $\Omega_{t}$ are the Rabi frequencies of the probe laser, coupling laser, signal field, and tuner field coupled states, respectively, and $\Omega^*$ denotes the complex conjugate. Also,
\begin{equation}
\begin{array}{rcl}
A&=&-2\Delta_p \\
B&=&-2(\Delta_p+\Delta_c)\\
C&=&-2(\Delta_p+\Delta_c+\Delta_{sig})\\
D&=&-2(\Delta_p+\Delta_c-\Delta_{t}),
\end{array}
\end{equation}
where $\Delta_p$, $\Delta_c$, $\Delta_{sig}$, and $\Delta_{t}$ are the detunings of the probe laser, couple laser, signal field, and tuner field, respectively, defined as
\begin{equation}
\Delta_{p,c,sig,t}=\omega_{p,c,sig,t}-\omega_{12,23,34,35} \,\,\, ,
\label{detuningeq_split}
\end{equation}
where $\omega_{12,23,34,35}$ are the on-resonance angular frequencies of transitions $\ket{1}$-$\ket{2}$, $\ket{2}$-$\ket{3}$, $\ket{3}$-$\ket{4}$, and $\ket{3}$-$\ket{5}$, respectively, and $\omega_{p,c,sig,t}$ are the angular frequencies of the probe, coupling, signal, and tuner fields, respectively. Notably in this scheme, $\Delta_t/2\pi=0~\mathrm{MHz}$ because the tuning field is locked to the $\ket{3}$-$\ket{5}$ transition.

\begin{widetext}
\begin{equation}
\boldsymbol{\cal{L}}=\begin{bmatrix}
\Gamma_2 \rho_{22} & -\gamma_{12}\rho_{12} & -\gamma_{13}\rho_{13} & -\gamma_{14}\rho_{14}& -\gamma_{15}\rho_{15}   \\

-\gamma_{21}\rho_{21} & \Gamma_3 \rho_{33} - \Gamma_2 \rho_{22} & -\gamma_{23}\rho_{23} & -\gamma_{24}\rho_{24}& -\gamma_{25}\rho_{25}  \\

-\gamma_{31}\rho_{31} & -\gamma_{32}\rho_{32} & \Gamma_{4}\rho_{44}+\Gamma_{5}\rho_{55}-\Gamma_{3}\rho_{33} & -\gamma_{34}\rho_{34}& -\gamma_{35}\rho_{35}  & \\

-\gamma_{41}\rho_{41} & -\gamma_{42}\rho_{42} & -\gamma_{43}\rho_{43} &  -\Gamma_4 \rho_{44} & -\gamma_{45}\rho_{45}  \\

-\gamma_{51}\rho_{51} & -\gamma_{52}\rho_{52} & -\gamma_{53}\rho_{53} & -\gamma_{45}\rho_{45}  &  -\Gamma_5 \rho_{55}
\end{bmatrix}
\label{L_split}
\end{equation}
\end{widetext}

For this system, the $\boldsymbol{\cal{L}}$ matrix is given in Eq.~(\ref{L_split}), where $\gamma_{ij}=(\Gamma_i+\Gamma_j)/2$ and $\Gamma_{i, j}$ are the transition decay rates. Since the purpose of the present study is to explore the intrinsic limitations of Rydberg-EIT field sensing in vapor cells, no collision terms or dephasing terms are added. While Rydberg-atom collisions, Penning ionization, and ion electric fields can, in principle, cause dephasing, such effects can, for instance, be alleviated by reducing the beam intensities, lowering the vapor pressure, or limiting the atom-field interaction time. In this analysis we set,
$\Gamma_1=0$, $\Gamma_2=2\pi\times$(6~{\rm MHz}),
$\Gamma_{3}=2\pi\times$(3~{\rm kHz}), and $\Gamma_{4,5}=2\pi\times$(2~{\rm kHz}).
Note, $\Gamma_{2}$ is for the D2 line in $^{133}$Cs \cite{SteckCsData}, and $\Gamma_{3,4,5}$ are typical Rydberg decay rates.

\subsection{Sequential Tuning}

For the sequential tuning scheme shown in Fig.~\ref{schematic}(d), the Hamiltonian can be expressed as:
\begin{equation}
\mathbf{H}=\frac{\hbar}{2}\begin{bmatrix}
    0 & \Omega_p & 0 & 0 & 0\\
    \Omega_p^* & A & \Omega_c & 0 & 0\\
    0 & \Omega_c^* & B & \Omega_{sig} & 0\\
    0 & 0 & \Omega_{sig}^* & C & \Omega_{t}\\
    0 & 0 & 0 & \Omega_{t}^* & D 
\end{bmatrix}\,\, ,
\label{H_seq}
\end{equation}
where $\Omega_p$, $\Omega_c$, $\Omega_{sig}$, and $\Omega_{t}$ are the Rabi frequencies of the probe laser, coupling laser, signal field, and tuner field coupled states, respectively. Also,

\begin{equation}
\begin{array}{rcl}
A&=&-2\Delta_p \\
B&=&-2(\Delta_p+\Delta_c)\\
C&=&-2(\Delta_p+\Delta_c+\Delta_{sig})\\
D&=&-2(\Delta_p+\Delta_c+\Delta_{sig}+\Delta_{t}),
\end{array}
\end{equation}
where $\Delta_p$, $\Delta_c$, $\Delta_{sig}$, and $\Delta_{t}$ are the detunings of the probe laser, couple laser, signal field, and tuner field, respectively, defined as
\begin{equation}
\Delta_{p,c,sig,t}=\omega_{p,c,sig,t}-\omega_{12,23,34,45} \,\,\, ,
\label{detuningeq_seq}
\end{equation}
where $\omega_{12,23,34,45}$ are the on-resonance angular frequencies of transitions $\ket{1}$-$\ket{2}$, $\ket{2}$-$\ket{3}$, $\ket{3}$-$\ket{4}$, and $\ket{4}$-$\ket{5}$ for the probe, coupling, signal, and tuner fields, respectively, and $\omega_{p,c,sig,t}$ are the angular frequencies of the probe, coupling, signal, and tuner fields, respectively. 

The $\boldsymbol{\cal{L}}$ matrix of this system is given in eq.~(\ref{L_seq}).

\begin{widetext}
\begin{equation}
\boldsymbol{\cal{L}}=\begin{bmatrix}
\Gamma_2 \rho_{22} & -\gamma_{12}\rho_{12} & -\gamma_{13}\rho_{13} & -\gamma_{14}\rho_{14}& -\gamma_{15}\rho_{15}   \\

-\gamma_{21}\rho_{21} & \Gamma_3 \rho_{33}-\Gamma_2 \rho_{22} & -\gamma_{23}\rho_{23} & -\gamma_{24}\rho_{24}& -\gamma_{25}\rho_{25}  \\

-\gamma_{31}\rho_{31} & -\gamma_{32}\rho_{32} & \Gamma_4 \rho_{44}-\Gamma_3 \rho_{33} & -\gamma_{34}\rho_{34}& -\gamma_{35}\rho_{35}  & \\

-\gamma_{41}\rho_{41} & -\gamma_{42}\rho_{42} & -\gamma_{43}\rho_{43} &  \Gamma_5 \rho_{55}-\Gamma_4 \rho_{44}& -\gamma_{45}\rho_{45}  \\

-\gamma_{51}\rho_{51} & -\gamma_{52}\rho_{52} & -\gamma_{53}\rho_{53} & -\gamma_{45}\rho_{45}  &  -\Gamma_5 \rho_{55}\\

\end{bmatrix}
\label{L_seq}
\end{equation}
\end{widetext}

Once again $\gamma_{ij}=(\Gamma_i+\Gamma_j)/2$ and $\Gamma_{i, j}$ are the transition decay rates, where 
$\Gamma_1=0$, $\Gamma_2=2\pi\times$(6~{\rm MHz}),
$\Gamma_{3}=2\pi\times$(3~{\rm kHz}), and $\Gamma_{4,5}=2\pi\times$(2~{\rm kHz}).

\subsection{Inverted Sequence}

For the inverted sequence scheme shown in Fig.~\ref{schematic}(e), the Hamiltonian can be expressed as:
\begin{equation}
\mathbf{H}=\frac{\hbar}{2}\begin{bmatrix}
0 & \Omega_p & 0 & 0 & 0 \\
\Omega_p^* & A & \Omega_c & 0 & 0 \\
0 & \Omega_c^* & B & \Omega_{t} & 0 \\
0 & 0 & \Omega_{t}^* & C &\Omega_{sig} \\
0 & 0 & 0 & \Omega_{sig}^* & D 
\end{bmatrix}\,\, ,
\label{H_inv}
\end{equation}
where $\Omega_p$, $\Omega_c$, $\Omega_t$, and $\Omega_{sig}$ are the Rabi frequencies of the probe laser, coupling laser, tuner field, and signal field coupled states, respectively. Also,
\begin{equation}
\begin{array}{rcl}
A&=&-2\Delta_p \\
B&=&-2(\Delta_p+\Delta_c)\\
C&=&-2(\Delta_p+\Delta_c-\Delta_{t})\\
D&=&-2(\Delta_p+\Delta_c-\Delta_{t}-\Delta_{sig}),
\end{array}
\end{equation}
where $\Delta_p$, $\Delta_c$, $\Delta_{t}$, and $\Delta_{sig}$ are the detunings of the probe laser, couple laser, tuner field, and signal field, respectively, defined as
\begin{equation}
\Delta_{p,c,t,sig}=\omega_{p,c,t,sig}-\omega_{12,23,34,45} \,\,\, ,
\label{detuningeq_inv}
\end{equation}
where $\omega_{12,23,34,45}$ are the on-resonance angular frequencies of transitions $\ket{1}$-$\ket{2}$, $\ket{2}$-$\ket{3}$, $\ket{3}$-$\ket{4}$, and $\ket{4}$-$\ket{5}$ for the probe, coupling, tuner, and signal fields, respectively, and $\omega_{p,c,t,sig}$ are the angular frequencies of the probe, coupling, tuner, and signal fields, respectively.

The $\boldsymbol{\cal{L}}$ matrix of this system is given in eq.~(\ref{L_inv}).

\begin{widetext}
\begin{equation}
\boldsymbol{\cal{L}}=\begin{bmatrix}
\Gamma_2 \rho_{22} & -\gamma_{12}\rho_{12} & -\gamma_{13}\rho_{13} & -\gamma_{14}\rho_{14}& -\gamma_{15}\rho_{15}   \\

-\gamma_{21}\rho_{21} & \Gamma_5\rho_{55}-\Gamma_2 \rho_{22} & -\gamma_{23}\rho_{23} & -\gamma_{24}\rho_{24}& -\gamma_{25}\rho_{25}  \\

-\gamma_{31}\rho_{31} & -\gamma_{32}\rho_{32} & -\Gamma_3 \rho_{33} & -\gamma_{34}\rho_{34}& -\gamma_{35}\rho_{35}  & \\

-\gamma_{41}\rho_{41} & -\gamma_{42}\rho_{42} & -\gamma_{43}\rho_{43} &  \Gamma_3 \rho_{33}-\Gamma_4 \rho_{44}& -\gamma_{45}\rho_{45}  \\

-\gamma_{51}\rho_{51} & -\gamma_{52}\rho_{52} & -\gamma_{53}\rho_{53} & -\gamma_{45}\rho_{45}  &  \Gamma_4 \rho_{44}-\Gamma_5 \rho_{55}\\

\end{bmatrix}
\label{L_inv}
\end{equation}
\end{widetext}

Again $\gamma_{ij}=(\Gamma_i+\Gamma_j)/2$ and $\Gamma_{i, j}$ are the transition decay rates, but this time 
$\Gamma_1=0$, $\Gamma_2=2\pi\times$(6~{\rm MHz}), $\Gamma_{3,4}=2\pi\times$(2~{\rm kHz}), and $\Gamma_{5}=2\pi\times$(3~{\rm kHz}).

\section{Simple three-level model}\label{sec:simple3}

Here we describe a simple three-level model to understand the EIT peak position in our experiments.
We consider three bare Rydberg states in the sequential tuning arrangement of Fig.~\ref{schematic}(d), $\ket{I}\equiv\ket{56D}$, $\ket{J}\equiv\ket{54F}$, and $\ket{K}\equiv\ket{57D}$, with two RF fields applied, one at frequency $\omega_{sig}$ detuned by $\Delta_{sig}$ from the $\ket{I}\to\ket{J}$ transition, and the other at frequency $\omega_t$ detuned by $\Delta_t$ from the $\ket{J}\to\ket{K}$ transition.

These bare states and fields produce three nearly degenerate energy levels. 
The first is defined by the atom in state $\ket{I}$ with $N_{sig}$ photons at frequency $\omega_{sig}$ and $N_t$ photons at frequency $\omega_t$.
The second state is defined by the atom in state $\ket{J}$ with $N_{sig}-1$ photons at frequency $\omega_{sig}$ and $N_t$ photons at frequency $\omega_t$.
The third state is defined by the atom in state $\ket{K}$ with $N_{sig}-1$ photons at frequency $\omega_{sig}$ and $N_t-1$ photons at frequency $\omega_t$.
These states can be labeled
\begin{equation}
\begin{array}{rcl}
\ket{i}&=&\ket{I,N_{sig},N_t}\\
\ket{j}&=&\ket{J,N_{sig}-1,N_t}\\
\ket{k}&=&\ket{K,N_{sig}-1,N_t-1},
\end{array}
\end{equation}
with energies
\begin{equation}
\begin{array}{rcl}
E_i &=& E_I + N_{sig} \hbar \omega_{sig} + N_t \hbar \omega_t\\
E_j &=& E_J + (N_{sig}-1) \hbar \omega_{sig} + N_t \hbar \omega_t\\
&=&E_A+\hbar \Delta_{sig}\\
E_k &=& E_K + (N_{sig}-1) \hbar \omega_{sig} + (N_t-1) \hbar \omega_t\\
&=&E_A+\hbar(\Delta_{sig}+\Delta_t).
\end{array}
\end{equation}
The Hamiltonian in the rotating wave approximation is then
\begin{equation}
\begin{array}{rcl}
H_{seq} = E_i &+&\hbar\Delta_{sig}\ket{j}\bra{j}-\hbar(\Delta_{sig}+\Delta_t)\ket{k}\bra{k}\\
&+&\frac{\hbar\Omega_{sig}}{2}(\ket{j}\bra{i}+\ket{i}\bra{j})\\
&+&\frac{\hbar\Omega_{t}}{2}(\ket{k}\bra{j}+\ket{j}\bra{k}),
\end{array}
\end{equation}
which can be rewritten as
\begin{equation}
H_{seq}=E_i+\hbar\begin{bmatrix}
0 & \Omega_{sig}/2 & 0\\
\Omega_{sig}/2 & -\Delta_{sig} & \Omega_{t}/2\\
0 & \Omega_t/2 & -\Delta_{sig}-\Delta_t
\end{bmatrix}.
\end{equation}
Similarly, we can write the Hamiltonian matrix for the inverted tuning as
\begin{equation}
H_{inv}=E_k+\hbar\begin{bmatrix}
\Delta_{sig}+\Delta_t & \Omega_{sig}/2 & 0\\
\Omega_{sig}/2 & -\Delta_{t} & \Omega_{t}/2\\
0 & \Omega_t/2 & 0
\end{bmatrix}.
\end{equation}
Considering the different arrangement of Rydberg states, $\ket{H}\equiv\ket{53F}$, $\ket{I}\equiv\ket{56D}$, and $\ket{J}\equiv\ket{54F}$, and applying the signal field along $\ket{I}\to\ket{J}$ and the tuner along $\ket{H}\to\ket{I}$ we can write the split tuning case as
\begin{equation}
H_{split}=E_j+\hbar\begin{bmatrix}
\Delta_{t} & \Omega_{t}/2 & 0\\
\Omega_{t}/2 & 0 & \Omega_{sig}/2\\
0 & \Omega_{sig}/2 & -\Delta_{sig}.
\end{bmatrix}
\end{equation}

Lastly we turn to the special case in the sequential tuning case where $\Delta_{sig}$~=~$-\Delta_t$, which we noted is the condition where the Raman EIT peak emerges. 
In this case three energy eigenvalues can be calculated
\begin{equation}
    E_{0,\pm}=E_i,E_i+\frac{\hbar}{2}\Big(-\Delta_{sig}\pm
    \sqrt{\Omega^2_{sig}+\Omega^2_t+\Delta^2_{sig}}\Big).
\end{equation}

Here, the $E_\pm$ solutions correspond to AT splitting with an effective Rabi frequency $\Omega^2=\Omega^2_{sig}+\Omega^2_t$.
However, the third peak, $E_0$, is fixed in frequency at the location of the main EIT peak and corresponds to the two-photon Raman peak we observe.

\section*{Acknowledgements}
This research was developed with funding from the Defense Advanced Research Projects Agency (DARPA).
The views, opinions and/or findings expressed are those of the author and should not be interpreted as representing the official views or policies of the Department of Defense or the U.S. Government.
A contribution of the US government, not subject to copyright in the United States.

\bibliography{tuning}

\end{document}